\RequirePackage{fix-cm}
\documentclass[twocolumn]{svjour3}          

\usepackage{amsmath, amssymb}    
\usepackage{graphicx}   
\graphicspath{ {./img/} }
\usepackage{cite}
\usepackage{verbatim}   
\usepackage{color}      
\usepackage{subfigure}  
\usepackage{hyperref}   
\usepackage{pstricks}
\usepackage{psfrag}
\newcommand{\ket}[1]{\ensuremath{|#1\rangle}}
\newcommand{\bra}[1]{\ensuremath{\langle #1|}}
\newcommand{\elemm}[3]{\ensuremath{\bra{#1}#2\ket{#3}}}
\newcommand{\braket}[2]{\ensuremath{\langle #1|#2\rangle}}

\journalname{Applied Physics B}
\begin{document}
\title{A study of one-dimensional transport of Bose-Einstein 
  condensates using exterior complex scaling}

\author{Julien Dujardin \and Alejandro Saenz \and Peter Schlagheck}

\institute{Julien Dujardin and Peter Schlagheck  \at
D\'{e}partement de Physique, Universit\'e de Li\`ege, 4000 Li\`ege, Belgium           \and
           Alejandro Saenz \at
Institut f\"{u}r Physik, Humboldt-Universit\"{a}t zu Berlin, 12489 Berlin, Germany
}

\date{Received: date / Revised version: date}

\maketitle



\begin{abstract}

We numerically investigate the one-dimen\-sional transport of 
Bose-Einstein condensates in the context of guided atom lasers
using a mean-field description of the condensate in terms of a 
spatially discretized Gross-Pitaevskii equation.
We specifically consider a waveguide configuration in which spatial
inhomogeneities and nonvanishing atom-atom interactions are restricted
to a spatially localized scattering region of finite extent.
We show how the method of smooth exterior complex scaling can be implemented
for this particular configuration in order to efficiently absorb the outgoing 
flux within the waveguide.
A numerical comparison with the introduction of a
complex absorbing potential as well as with the analytically exact
elimination of the dynamics of the free non-interacting motion outside
the scattering region, giving rise to
transparent boundary conditions, clearly confirms the accuracy and 
efficiency of the smooth exterior complex scaling method.

\end{abstract}
\section{Introduction}
\label{sec:introduction}
The perspective to realize atomtronic devices \cite{MicO04PRL,SeaO07PRA,PepO09PRL} 
as well as the exploration of transport features that are known from electronic 
mesoscopic systems \cite{BraO12S} have strongly stimulated the research on the 
dynamical properties of ultracold atoms in open systems.
While fermionic atoms provide direct analogies with the electronic case 
\cite{BraO12S,BruBel12PRA,KriO13PRL}, the use of a bosonic atomic species 
brings along new aspects and challenges for the atomic transport problem 
\cite{GutGefMir12PRB,Iva13EPJB}.
It is in this context particularly relevant to quantitatively 
understand the conceptual link between a mesoscopic Bose-Einstein
condensate, which may serve as a reservoir for a bosonic transport
setting, and the microscopic dynamics of an ensemble of few
interacting atoms that encounter each other \textit{e.g.} within a 
transistor-like device.
A particularly promising configuration for the experimental study of
these latter aspects is provided by the guided atom laser
\cite{Guerin2006,Couvert2008a,Gattobigio2011} in which atoms are
coherently outcoupled from a trapped Bose-Einstein condensate into an
optical waveguide.
A coherent atomic beam can thereby be created and injected onto engineered
optical scattering geometries, which would allow one to study bosonic
many-body scattering at well-defined incident energy.

A theoretical modelling of such waveguide scattering processes within guided
atom lasers faces the problem of dealing with an open system in a many-body
context.
Within the framework of the mean-field approximation  described by the nonlinear
Gross-Pitaevskii equation, this problem can be solved to a satisfactory degree
by imposing absorbing boundary conditions \cite{Shi91PRB} at the two open ends of 
the numerical grid representing the waveguide, which are suitably defined 
in order to match the dispersion relation of the expected outgoing waves 
\cite{PauRicSch05PRL,PauO07PRA}.
While this approach provides a reasonably efficient absorption of outgoing
Gross-Pitaevskii waves even in the presence of dynamical instabilities
\cite{PauO05PRA}, it ultimately breaks down if quantum fluctuations beyond
the Gross-Pitaevskii approximation are taken into account in the theoretical
description of the bosonic scattering process \cite{ErnPauSch10PRA}.
Complex absorbing potentials (CAPs) \cite{Kosloff1986,Riss1998} that exhibit a nonvanishing imaginary part 
can still be introduced in that case in order to damp the outgoing flux.
However, their numerical implementation requires great care in order to suppress 
unwanted backreflections of outgoing waves at the onset of the artificially
introduced imaginary potential (see Refs.~\cite{MoiO04JPB,RapO10PRA} for
successful applications of CAPs in the context of the Gross-Pitaevskii equation).

The method of \emph{Complex Scaling} (CS)
\cite{Balslev1971,BarrySimon1973,Reinhardt1982,Junker1982,Ho1983,Lowdin1988,Moiseyev1998}
provides a more satisfactory alternative from a conceptual point of view.
This method essentially consists (in 1D) in the rotation 
$x \mapsto z = x \exp(i\theta)$ of the spatial coordinate in the 
complex plane by a suitably chosen angle $\theta>0$.
Decaying quasi-bound states that exhibit outgoing boundary conditions become
square integrable by this transformation and can thereby be computed in an open system.
The complex scaling approach can formally be generalized to the nonlinear
Gross-Pitaevskii equation \cite{MoiCed05PRA,SchPau06PRA}.
However, its practical implementation in this latter context poses substantial
numerical difficulties concerning the proper evaluation of the nonlinear term
in the complex rotated frame \cite{SchPau06PRA,WimSchMan06JPB,SchWim07APB}.

In this paper, we focus on the method of 
\emph{Exterior Complex Scaling} (ECS) \cite{Simon1979} which is 
particularly suited for open systems in which potential scattering and 
(mean-field) particle-particle interaction effects are restricted to a finite 
spatial region.
This method consists in a complex rotation of the position 
coordinate applied only to the asymptotic spatial domain of freely 
outgoing and noninteracting particles. 
ECS has been applied in a wide range of problems such as 
computing the probability distribution of excitations to the electronic 
continuum of HeT$^{+}$ following the $\beta$ decay of the T$_2$ molecule 
\cite{Froelich1993}, molecular photoionization \cite{Saenz2003}, 
electron-hydrogen collisions \cite{Rescigno1999,McCurdy2004a} and 
also strong-field infrared photo-ionization of atoms 
\cite{He2007, Tao2009,Tao2012,Scrinzi2012}. 
While this approach exactly reproduces the true decay behaviour in the 
open quantum system from a formal point of view (in contrast to the 
introduction of CAPs), numerical imprecisions are necessarily introduced 
through the discretization of space in the finite-difference approximation 
\cite{McCurdy2004a} of the Gross-Pitaevskii approximation. 
This problem can be overcome by using high rank finite elements 
\cite{Scrinzi1993,Scrinzi2010} or a B-spline basis \cite{McCurdy2004} 
instead of a finite-difference representation. 
Alternatively, an analytic transition function can be used to interpolate 
from the scaled ``outer'' domain to the unscaled ``inner'' domain which 
may contain all sorts of nontrivial scattering and interaction phenomena. 
This latter method is named \emph{Smooth Exterior Complex Scaling} 
(SECS) \cite{Rom1990}. 
It has been used, for example, to compute doubly excited states 
of the helium atom \cite{Elander2003}
and to investigate the dynamical stability of stationary scattering
states of a Bose-Einstein condensate in two-dimensional billiard geometries
\cite{HarO12AP}.

The main aim of this study is to 
assess the applicability of smooth exterior complex scaling 
to the mean-field transport of Bose-Einstein condensates in
one-dimensional waveguides using the finite-difference approximation.
We therefore represent, as described in 
Sec.~\ref{sec:DIPBH}, the waveguide by means 
of a discrete one-di\-mensional chain which is at some point connected to a 
separate site representing the reservoir trap of the atom laser.
Scattering and interaction phenomena are assumed to be restricted to a 
finite spatial domain within this chain.
As is shown in Sec.~\ref{sec:OBH}, this crucial assumption allows us 
to formally separate this central domain from the two attached semi-infinite 
``leads'' featuring free non-interacting motion.
This gives rise to perfectly transparent boundary conditions which render, 
however, the numerical propagation of the system rather time-consuming.
In Sec.~\ref{sec:SECS}, smooth exterior complex scaling is then introduced 
to this open system as a feasible alternative.
Finally, numerical results comparing the use of smooth exterior complex 
scaling, of complex absorbing potentials, as well as of the transparent 
boundary conditions derived in Sec.~\ref{sec:OBH} are presented in 
Sec.~\ref{sec:Comp}.

\section{1D Bose-Hubbard chain with a source}
\label{sec:DIPBH}
We consider an infinite one-dimensional (1D) Bose-Hub\-bard (BH) 
system representing the transverse ground mode of a 1D waveguide 
in a finite-difference representation.
This BH chain is connected at one of its sites to
one additional site representing a reservoir of 
Bose-Einstein condensed atoms with the chemical potential $\mu$,
as illustrated in Fig.~\ref{fig:BHproj}.
The many-body Hamiltonian of this system reads
\begin{eqnarray}
\label{eq:BHham}
	\hat{\mathcal{H}} &= \displaystyle\sum_{\ell=-\infty}^{+\infty}& \Bigg[ -J(\hat{a}^\dagger_{\ell+1}\hat{a}_{\ell} + \hat{a}^\dagger_{\ell}\hat{a}_{\ell+1}) \nonumber \\
				&&+ \frac{g_\ell}{2} \hat{n}_\ell(\hat{n}_\ell-1) + V_\ell\hat{n}_\ell \Bigg] \nonumber \\ 
				&&+ \kappa^*(t)\hat{b}^\dagger\hat{a}_{\ell_S} + \kappa(t)\hat{a}^\dagger_{\ell_S}\hat{b} + \mu \hat{b}^\dagger \hat{b},
\end{eqnarray}
where we define by $\hat{a}_\ell$ and $\hat{a}_\ell^\dagger$ the 
annihilation and creation operators, respectively, on the site $\ell$ 
of the chain, with $\hat{n}_\ell = \hat{a}^\dagger_\ell\hat{a}_{\ell}$ 
the corresponding number operator,
and by $\hat{b}$ and $\hat{b}^\dagger$ the annihilation and creation 
operator of the reservoir to which the chain is connected at the site $\ell_S$. 
The hopping strength $J$, the on-site interaction strength $g_\ell$,
as well as the on-site potential $V_\ell$ can be determined from the
Hamiltonian of the underlying continous system through the discretization 
of the spatial coordinate along the waveguide.
The coupling strength $\kappa(t)$, on the other hand, is related to the 
outcoupling process of atoms from the reservoir and can be controlled
in a time-dependent manner (\textit{e.g.} through the variation of the
intensity of a radiofrequency field in the case of 
Refs.~\cite{Guerin2006,RioO08PRA}).
We should mention, however, that the framework developed here does not
exclusively apply to guided atom lasers, but could also be used in the
context of analogous transport processes taking place within optical lattices.

In the Heisenberg representation, the time evolution of the annihilation 
operators $\hat{a}\equiv\hat{a}_\ell(t)$ and $\hat{b}\equiv\hat{b}(t)$
is given by the Heisenberg equations (we set $\hbar=1$ in the following)
\begin{subequations}
\label{eq:inffieldopevol}
 \begin{align}
	i \frac{\partial \hat{a}_\ell(t)}{\partial t} =& V_\ell\hat{a}_\ell(t) - J\left[\hat{a}_{\ell-1}(t) + \hat{a}_{\ell+1}(t)\right] \nonumber\\
	&  + g_\ell \hat{a}_\ell^\dagger(t)\hat{a}_\ell(t)\psi_\ell(t) + \kappa(t)\delta_{\ell,\ell_S}\hat{b}(t)\\
	i \frac{\partial\hat{b}(t)}{\partial t} =& \mu\hat{b}(t) + \kappa^*(t)\hat{a}_{\ell_S}(t).
 \end{align}
\end{subequations}
In accordance with the working principle of an atom laser, we 
consider an initial state at time $t_0$ in which 
the source is populated with a very large number $N\to\infty$ of atoms 
and the chain is empty. 
Moreover, we consider a very weak coupling strength 
$\kappa(t)\to 0$ which tends to zero such that $N|\kappa(t)|^2$ remains finite.
This combined limit gives rise to a finite population within the chain,
which is (at time-independent $\kappa$) alimented by a steady flux of atoms
from the reservoir.

In the following, we consider the classical mean-field regime 
of large on-site densities and weak interaction strengths within the 
Bose-Hubbard chain, in which (at finite evolution time $t$) the system can 
be described using c-numbers instead of operators.
The dynamics of the system in this regime is described by the
nonlinear Gross-Pitaevskii (GP) equation
\begin{subequations}
\label{eq:BHsystem}
 \begin{align}
 \label{eq:chain}
i\frac{\partial \psi_\ell(t)}{\partial t} &= (V_\ell-\mu)\psi_\ell(t)
							 -J\left(\psi_{\ell+1}(t) + \psi_{\ell-1}(t)\right) \nonumber\\
							 &  + g_\ell|\psi_\ell(t)|^2 \psi_\ell(t)+ \delta_{\ell,\ell_S}\kappa(t)\chi(t) \\
\label{eq:Source}
i\frac{\partial \chi(t)}{\partial t} &=  \kappa^*(t)\psi_{\ell_S}(t)
 \end{align}
\end{subequations}
where we define the amplitudes 
$\psi_\ell(t)=\langle\hat{a}_\ell(t)\rangle e^{-i\mu t}$ 
and $\chi(t)=\langle\hat{b}(t)\rangle e^{-i\mu t}$
with the initial conditions $\psi_\ell(t_0)=0$ and $\chi(t_0)=\sqrt{N}$.
From Eqs.~\eqref{eq:chain} and \eqref{eq:Source} we can infer that
$\chi(t) = \sqrt{\mathcal{N}}(1+\mathcal{O}(|\kappa|^2)$ at finite $t-t_0$, 
and as a consequence we can neglect the time dependence of $\chi$
in the limit $\kappa\to 0$. 
We then obtain a discrete nonlinear Schr\"{o}dinger equation with a 
source term:
\begin{eqnarray}
 \label{eq:chain2}
i\frac{\partial \psi_\ell(t)}{\partial t} &=& (V_\ell-\mu)\psi_\ell(t)  \nonumber\\
							 &&-J\left(\psi_{\ell+1}(t) + \psi_{\ell-1}(t)\right) \nonumber\\
							 &&  + g_\ell|\psi_\ell(t)|^2 \psi_\ell(t)+\delta_{\ell,\ell_S} \kappa(t)\sqrt{\mathcal{N}}\,.
\end{eqnarray}

\section{Transparent boundary conditions}
\label{sec:OBH}

The standard procedure for a numerical study of the time-dependent 
dynamics within an infinite chain consists in defining a sufficiantly large 
``simulation box'' containing a finite number of sites.
The choice of the boundary conditions at the edges of the box is, in 
general, irrelevant for wave packet evolution processes that evolve
within a finite time; it does, however, matter for the type of
scattering processes that we consider here:
choosing hard-wall or periodic boundary conditions would rather quickly 
lead to unwanted reflections of the matter wave at the artificially 
introduced boundaries of the box.

\begin{figure}[t]
\centering
    \begin{psfrags}
      \psfrag{S}{{\small Source}}
      \psfrag{0}{{\small $0$}}
      \psfrag{1}{{\small $1$}}
      \psfrag{N}{{\small $L$}}
      \psfrag{N+1}{{\small $L+1$}}
      \psfrag{nS}{{\small $\ell_S$}}
      \psfrag{L}{{\small $\mathcal{L}$}}
      \psfrag{Q}{{\small $\mathcal{Q}$}}
      \psfrag{R}{{\small $\mathcal{R}$}}      
      \includegraphics[width=0.8\linewidth]{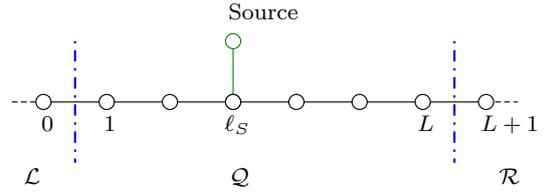}
    \end{psfrags}
	\caption{(color online) One-dimensional infinite BH system with an additional site for the source. The zone $\mathcal{Q}$ is defined between the dot-dashed lines. \label{fig:BHproj}}
\end{figure}

To avoid such artificial backreflections, we can introduce 
\emph{transparent boundary conditions} (TBC), making use of the fact, as
explained in the introduction, that the scattering potential and the 
interaction strength are non-zero only in a finite region of space. 
To this end, we formally divide the system in three parts, namely 
the semi-infinite left and right parts $\mathcal{L}$ and $\mathcal{R}$ 
where neither interaction nor scattering takes place, and the finite central
part $\mathcal{Q}$ consisting of $L$ sites numbered from $1$ to $L$,
which contains potential scattering, atom-atom interaction, as well
as the link to the source (see Fig.~\ref{fig:BHproj}). 
For the sake of compactness of the formalism, we regroup all the 
amplitudes $\psi_\ell$ into a state $\ket{\psi}$ defined through
\begin{equation}
\ket{\psi} = \sum_{\ell=-\infty}^{+\infty} \psi_\ell \ket{\ell},
\end{equation}
where the on-site states $\ket{\ell}$ form an orthonormal basis 
$\braket{\ell}{\ell'}=\delta_{\ell,\ell'}$. 
Formally Eq.~\eqref{eq:chain2} can then be expressed as 
\begin{equation}
i \frac{\partial \ket{\psi}}{\partial t} = \mathcal{H}\ket{\psi} = [\mathcal{H}_f+\mathcal{V}+\mathcal{U}(\psi)]\ket{\psi} + \ket{S}
\end{equation}
with $\ket{S} = \kappa(t)\sqrt{\mathcal{N}} \ket{\ell_0}$,
where we decompose the Gross-Pitaveksii Hamiltonian $\mathcal{H}$ in 
the free motion on the chain described by $\mathcal{H}_f$, the scattering 
potential included in $\mathcal{V}$, and the nonlinear interaction term 
$\mathcal{U}(\psi)$. 
The corresponding matrix elements in the local basis are
\begin{subequations}
\begin{align}
\label{eq:kinelements}
\elemm{\ell}{\mathcal{H}_f}{\ell'} &= -\mu \delta_{\ell,\ell'} - J (\delta_{\ell,\ell'-1} +\delta_{\ell,\ell'+1}),\\
\elemm{\ell}{\mathcal{V}}{\ell'} &= \delta_{\ell,\ell'} V_{\ell},\\
\elemm{\ell}{\mathcal{U}(\psi)}{\ell'} &= \delta_{\ell,\ell'} g_\ell |\psi_\ell|^2.
\end{align}
\end{subequations}

We can now define the division of the system described above 
using the Feshbach projection formalism with the three projectors
\begin{subequations}
\begin{align}
P_\mathcal{L} &= \sum_{\ell=-\infty}^0 \ket{\ell}\bra{\ell}, \\
P_\mathcal{Q} &= \sum_{\ell=1}^L \ket{\ell}\bra{\ell}, \\
P_\mathcal{R} &= \sum_{\ell=L+1}^\infty \ket{\ell}\bra{\ell}.
\end{align}
\end{subequations}
This gives rise to the three coupled evolution equations 
\begin{subequations}
\begin{align}
i\frac{\partial \ket{\psi^{(\mathcal{L})}}}{\partial t} &= \mathcal{H}_\mathcal{L}\ket{\psi^{(\mathcal{L})}}+ \mathcal{W}_{\mathcal{LQ}}\ket{\psi^{(\mathcal{Q})}}
\label{eq:left} ,\\
i\frac{\partial \ket{\psi^{(\mathcal{Q})}}}{\partial t}  &=(\mathcal{H}_\mathcal{Q} + \mathcal{V}_{\mathcal{Q}} + \mathcal{U_{\mathcal{Q}}}(\psi))\ket{\psi^{(\mathcal{Q})}} + \ket{S} \nonumber\\
 &  +\mathcal{W}_{\mathcal{QL}}\ket{\psi^{(\mathcal{L})}} +\mathcal{W}_{\mathcal{QR}}\ket{\psi^{(\mathcal{R})}}, 
\label{eq:center} \\
i\frac{\partial \ket{\psi^{(\mathcal{R})}}}{\partial t} &= \mathcal{H}_\mathcal{R}\ket{\psi^{(\mathcal{R})}}+ \mathcal{W}_{\mathcal{RQ}}\ket{\psi^{(\mathcal{Q})}}
\label{eq:right}
\end{align}
\end{subequations}
where we define $\ket{\psi^{(\mathcal{X})}} = P_\mathcal{X}\ket{\psi}$,
$\mathcal{H}_\mathcal{X} = P_\mathcal{X}\mathcal{H}_fP_\mathcal{X}$,
with $\mathcal{X}$, $\mathcal{Y}$ being equal to 
$\mathcal{Q}, \mathcal{L}$ or $\mathcal{R}$, as well as
\begin{eqnarray}
\mathcal{W}_{\mathcal{LQ}} &=& P_\mathcal{L}\mathcal{H}_fP_\mathcal{Q} =
-J \ket{0}\bra{1} = \mathcal{W}_{\mathcal{QL}}^\dagger,\\
\mathcal{W}_{\mathcal{RQ}} &=& P_\mathcal{R}\mathcal{H}_fP_\mathcal{Q} =
-J \ket{L+1}\bra{L} = \mathcal{W}_{\mathcal{QR}}^\dagger.
\end{eqnarray}
The evolution equations \eqref{eq:left}, \eqref{eq:right} for 
the left and the right part are linear and describe a free propagation 
in a semi-infinite lead.  
As a consequence, we can formally integrate them and plug the 
result in the evolution equation \eqref{eq:center} for the central part. 
This yields
\begin{eqnarray}
\label{eq:evol}
i\frac{\partial \ket{\psi^{(\mathcal{Q})}}}{\partial t}  &=&[\mathcal{H}_\mathcal{Q}+ \mathcal{V}_\mathcal{Q} + \mathcal{U}_\mathcal{Q}(\psi)]\ket{\psi^{(\mathcal{Q})}} \nonumber\\
 & & -i\int_{t_0}^t dt'\mathcal{W}_{\mathcal{QL}}e^{-i(t-t')\mathcal{H}_\mathcal{L}}\mathcal{W}_{\mathcal{LQ}}\ket{\psi^{(\mathcal{Q})}(t')}\nonumber\\
  & & -i\int_{t_0}^t dt'\mathcal{W}_{\mathcal{QR}}e^{-i(t-t')\mathcal{H}_\mathcal{R}}\mathcal{W}_{\mathcal{RQ}}\ket{\psi^{(\mathcal{Q})}(t')}\nonumber\\
  & & + \mathcal{W}_{\mathcal{QR}}e^{-i(t-t_0)\mathcal{H}_\mathcal{R}}\ket{\psi^{(\mathcal{R})}(t_0)}\nonumber\\
  & & + \mathcal{W}_{\mathcal{QL}}e^{-i(t-t_0)\mathcal{H}_\mathcal{L}}\ket{\psi^{(\mathcal{L})}(t_0)},
\end{eqnarray}
where the second and third lines describe the decay into the leads 
and the fourth and fifth lines describe the propagation of the initial 
conditions within the lead into the scattering region $\mathcal{Q}$.

The integrals in Eq.~\eqref{eq:evol} are calculated using the normalized 
continuum eigenstates $\ket{k^{\mathcal{(L/R)}}}$
of the leads, which in the local basis $\ket{\ell}$ can be written as
\begin{equation}
 \braket{\ell}{k^\mathcal{(L)}} = \sqrt{\frac{2}{\pi}}\sin[(\ell-1)k] 
 \quad \mbox{with}\; \ell< 1
\end{equation}
for the left lead and
\begin{equation}
 \braket{\ell}{k^\mathcal{(R)}} = \sqrt{\frac{2}{\pi}}\sin[(\ell-L)k]
 \quad \mbox{with}\; \ell> L
\end{equation}
for the right lead, with $0\le k \le \pi$,
$\braket{k^{\mathcal{(L/R)}}}{{\tilde{k}^{\mathcal{(L/R)}}}} = \delta(k-\tilde{k})$
and the associated eigenvalues
\begin{equation}
 E_k = -2J\cos(k)-\mu.
\end{equation}
For the term $\mathcal{W}_{\mathcal{QL}}e^{-i\tau\mathcal{H}_\mathcal{L}}
\mathcal{W}_{\mathcal{LQ}}$ for instance, we obtain the expression
\begin{equation*}
\mathcal{W}_{\mathcal{QL}}e^{-i\tau\mathcal{H}_\mathcal{L}}\mathcal{W}_{\mathcal{LQ}} = J^2\int_0^\pi dk \,|\braket{0}{k^{(\mathcal{L}})}|^2 e^{-i\tau E_k}\ket{1}\bra{1}
\end{equation*}
which is related to Bessel integrals. 
This finally yields a finite set of $L$ integro-differential 
equations
\begin{eqnarray}
\label{eq:integro_evol}
i \frac{\partial \psi_\ell}{\partial t} &=& (V_\ell-\mu)\psi_\ell - J(\psi_{\ell-1}\theta_{\ell-1,1}+ \psi_{\ell+1}\theta_{L,\ell+1}) \nonumber\\
	& & + g_\ell |\psi_\ell|^2\psi_\ell + \kappa(t)\sqrt{N}\delta_{\ell,\ell_S} \nonumber\\
	& & - 2i(\delta_{\ell,1} + \delta_{\ell,L})J^2 \int_{t_0}^t dt'\, \mathcal{M}_{1}(t-t')\psi_\ell(t') \nonumber\\
	& & +2J \delta_{\ell,1}\sum_{\ell'=-\infty}^0 \mathcal{M}_{\ell'-1}(t-t_0)\psi_{\ell'}(t_0) \nonumber\\
	& & -2J \delta_{\ell,L}\sum_{\ell'=L+1}^\infty \mathcal{M}_{\ell'-L}(t-t_0)\psi_{\ell'}(t_0)
\end{eqnarray}
with
\begin{equation}
\theta_{\ell,\ell'}=\left\{\begin{matrix} 1 &  & \mathrm{if}\; \ell\ge\ell' \\ 0 &  & \mathrm{otherwise} \end{matrix}\right.
\end{equation}
and
\begin{equation}
\mathcal{M}_\ell(\tau) = \frac{i^\ell}{2} \left[ J_{\ell-1}
\left(2J\tau\right)+J_{\ell+1}\left(2J\tau\right)\right]e^{i\mu\tau}
\end{equation}
where $J_\ell(x)$ is the Bessel function of the first kind.
As no approximation has been made in this section, 
Eq.~\eqref{eq:integro_evol} reproduces the true evolution of the infinite 
system under consideration, described by Eq.~\eqref{eq:chain2}.
The integral term in the third line of Eq.~\eqref{eq:integro_evol} describing
the decay into the left and right leads therefore yields a perfectly
transparent boundary condition that is defined on the first and last site
of the central region.

\section{Smooth Exterior Complex Scaling}
\label{sec:SECS}
Within a continuous 1D system, the method of complex scaling 
consists in the transformation
$x \mapsto z = x e^{i\theta}$ 
($x\in\mathbb{R}$) of the position coordinate with $\theta > 0$.
With this transformation, a stationary wave $\psi(x) \sim e^{ikx}$ 
becomes $\psi(z) \sim e^{ikz}=e^{ikx\cos \theta}e^{-kx\sin \theta}$ 
where $k$ is the wavenumber. 
Waves traveling from left to the right ($k>0$) are therefore subject 
to damping for positive $x$, while waves traveling from right to the 
left ($k<0$) are damped for negative $x$ (and would be enhanced for 
positive $x$ \cite{Scrinzi2010}). 
Thus complex scaling allows to describe in a numerically efficient manner
the outgoing waves that arise in our 1D scattering problem in which the 
source is part of the scattering system (see Fig.~\ref{fig:BHproj}).

For our case, we want to apply the complex scaling transformation to the
leads $\mathcal{L}$ and $\mathcal{R}$, while the finite scattering
region $\mathcal{Q}$ is supposed to remain unscaled. 
In order to properly introduce the method of smooth exterior complex scaling
(SECS) for this case, we first consider a continuous system
with a wavefunction $\psi(x,t)$ that is subject to the Schr\"odinger
equation
\begin{equation}
i\frac{\partial}{\partial t} \psi(x,t) = -J\frac{\partial^2}{\partial x^2}
\psi(x,t) \,.
\end{equation}
We now define a complex analytical function $q(x)$ on the 1D space and 
introduce an (in general non-unitary) transformation $\mathcal{U}$ through
\begin{equation}
  \mathcal{U} \psi(x,t) = \psi(z(x), t)
\end{equation}
where $z(x)$ is defined as 
\begin{equation}
z(x) = \int_{0}^x q(x')dx'
\end{equation}
(assuming, without loss of generality, that the spatial origin 
$x=0$ is part of $\mathcal{Q}$).
The evolution of the transformed wavefunction is then given by
\begin{equation}
  i\frac{\partial \mathcal{U} \psi}{\partial t}(x,t) =
  -\frac{J}{q^2(x)} \left( \frac{\partial^2}{\partial x^2}
  - \frac{q'(x)}{q(x)} \frac{\partial}{\partial x} \right)
  \mathcal{U} \psi(x,t) \label{eq:kincompscal}
\end{equation}
where $q'(x)$ is the first derivative of $q$ with respect to $x$. 

The goal is to choose $q(x)$ such that the Hamiltonian remains 
unscaled in the $\mathcal{Q}$ region and scaled in the other two regions. 
For this purpose, we choose $q(x)~\to~1$ within $\mathcal{Q}$ and 
smoothly ramp $q(x)$ to $e^{i \theta}$ within the scaled regions.
The function $q(x)$ we used in this study reads \cite{Kalita2011}
\begin{equation}
q(x) = 1+(e^{i\theta}-1)\left(1+\frac{f_{+}(x)-f_{-}(x)}{2}\right)
\label{eq:qx}
\end{equation}
with
\begin{equation}
  f_{\pm}(x) = \tanh(\lambda(x-x_\pm)\pm 2\pi)
\end{equation}
where $\lambda$ is defined as the smoothing parameter and the interval 
$[x_{-},x_{+}]$ corresponds to the $\mathcal{Q}$ region. 

In order to apply SECS to our BH chain, which can be seen as a 
discretization of space, we need to define the 
matrix elements of the spatial derivatives appearing in 
Eq.~(\ref{eq:kincompscal}) within the discrete basis of on-site
states $\ket{\ell}$.
Within the framework of the finite-difference approximation, we find
\begin{subequations}
\begin{align}
\label{eq:scaledkinelements}
\elemm{\ell}{\frac{\partial}{\partial x }}{\ell'} &= \frac{1}{2} (\delta_{\ell,\ell'+1} -\delta_{\ell,\ell'-1}),\\
\elemm{\ell}{\frac{\partial^2}{\partial x^2 }}{\ell'} &= \delta_{\ell,\ell'+1} + \delta_{\ell,\ell'-1} - 2 \delta_{\ell,\ell'},
\end{align}
\end{subequations}
which yields, using Eq.~\eqref{eq:kinelements}, the relation
\begin{equation}
  -J \elemm{\ell}{\frac{\partial^2}{\partial x^2 }}{\ell'} =
  \elemm{\ell}{\hat{\mathcal{H}}_f}{\ell'} + (\mu-2J)\delta_{\ell,\ell'}
\end{equation}
between the free 1D kinetic energy and the hopping term of the BH model.
Defining $\bra{\ell'}q\ket{\ell}=q_\ell\delta_{\ell,\ell'}$, we can 
discretize Eq~\eqref{eq:kincompscal}.
Provided that the transition between the scaled and unscaled regions
is sufficiently smooth (\textit{i.e.} $\lambda\ll 1$), we can set 
$q_{\ell-1}\simeq q_\ell\simeq q_{\ell+1}$ 
and the evolution equation \eqref{eq:integro_evol} now reads
\begin{eqnarray}
\label{eq:SECS}
i\frac{\partial \psi_\ell}{\partial t} &=& \left(V_\ell- \mu q_\ell\right)\psi_\ell + g_\ell|\psi_\ell|^2 \psi_\ell + \kappa(t)\sqrt{\mathcal{N}}\delta_{\ell,\ell_S} \nonumber \\
    & & +2J( q_\ell + q_\ell^{-1}) \psi_\ell \nonumber \\
	& & -J\left[\frac{1}{q_{\ell+1}} +\frac{1}{2}\frac{q'_{\ell+1}}{ q^2_{\ell+1}}\right] \psi_{\ell+1} \nonumber \\
	& & -J \left[\frac{1}{q_{\ell-1}} -\frac{1}{2}\frac{q'_{\ell-1}}{ q^2_{\ell-1}}\right]\psi_{\ell-1} \nonumber \\
	& & +2J \delta_{\ell,1}\sum_{\ell'=-\infty}^0 \mathcal{M}_{\ell'-1}(t-t_0)\psi_{\ell'}(t_0) \nonumber\\
	& & -2J \delta_{\ell,L}\sum_{\ell'=L+1}^\infty \mathcal{M}_{\ell'-L}(t-t_0)\psi_{\ell'}(t_0).
\end{eqnarray}
The index $\ell$ can now take values in $\mathbb{Z}$ and the unscaled 
region goes from $\ell_{-}=1$ to $\ell_{+}=L$. 
In the practical numerical implementation, the BH chain has to be 
sufficiently long in order to absorb the outgoing flux.
The last two lines of Eq.~\eqref{eq:SECS} still contain the propagation 
of the initial population of the leads into the scattering region, 
which is unaffected by the SECS transformation.

\section{Results}
\label{sec:Comp}
We now compare the SECS and TBC methods with each other and
with the well established method of complex absorbing potentials (CAPs). 
The imaginary part of such a complex potential renders the 
Hamiltonian non-Hermitian and thus the evolution non-unitary. 
For the sake of comparability with the SECS method, we choose the 
absorbing potential
\begin{equation}
 V^\textrm{CAP}_\ell = -i\,\, \textrm{Im}( q_\ell ),
\end{equation}
with $q_\ell$ being defined, as in Sec.~\ref{sec:SECS}, by the 
discretization of Eq.~\eqref{eq:qx}.

The TBC, SECS, and CAP methods are applied to the case of free kinetic 
propagation along a homogeneous and noninteracting BH chain, as well as to
the case of propagation across a symmetric double-barrier configuration 
in the presence of interaction, which can be seen as an atomic quantum dot.
At the initial time $t_0$, the BH chain is considered to be either completely
empty (\textit{i.e.} $\psi_\ell(t_0) = 0$ for all $\ell\in\mathbb{Z}$)
or populated with some randomly selected complex amplitudes on each site.
The coupling to the source is ramped according to
\begin{equation}
 \kappa(t) = \frac{1}{1+ e^{-(J(t-t_0) - 50) / 5 }}.
\end{equation}
To solve the differential equations, we use a Runge-Kutta Fehlberg (RKF) 
method.
This method is of order $\mathcal{O}(\delta t^4)$ with an error 
estimator of order $\mathcal{O}(\delta t^5)$, which allows one
to adapt the numerical time step $\delta t$ in order to keep the 
numerical solution as close as possible to the mathematically 
true solution of the equations.

\subsection{Free case with empty leads}
\label{subsec:free}
We first consider the case of a free kinetic propagation,
\textit{i.e.} $V_\ell=g_\ell=0$ for all $\ell\in\mathbb{Z}$, and 
compare the density profiles obtained by the TBC, SECS, and CAP methods 
against the expected value for the stationary density $n^\varnothing$.
The latter can be analytically calculated with Eq.~\eqref{eq:integro_evol} 
by restricting the central region $\mathcal{Q}$ to a single site.
We obtain
\begin{equation}
\label{eq:freen}
n^{\varnothing} = \lim_{t\to\infty} |\psi(t)|^2 = 
\lim_{t\to\infty}\frac{\mathcal{N}|\kappa(t)|^2}{4J^2-\mu^2}.
\end{equation}

\begin{figure}[t]
\begin{center}
\input{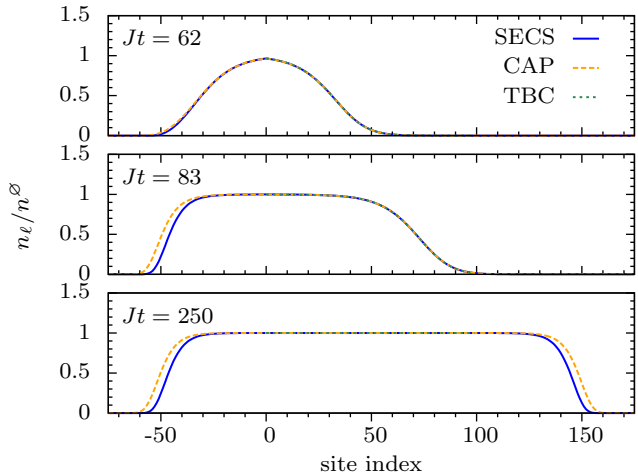}
\caption{(color online) Density profiles for the case of free propagation 
(\textit{i.e.} $g_\ell=V_\ell=0$) in a homogeneous BH chain 
for three different propagation times $t$.
The source of atoms is connected at site $\ell_S=1$. 
The TBC method (red solid line), the SECS method (green dashed line),
as well as the CAP method (blue dotted line) yield practically identical
densities within the region $\mathcal{Q}$ defined from site 1 to site 
100, which agree there with the analytical expression \eqref{eq:freen} for 
the stationary density.
Differences between the numerical methods naturally appear within
the leads (in which the results obtained by the TBC method are not 
displayed). 
While SECS is found to absorb the outgoing flux most efficiently 
in terms of computation time, TBC is the slowest method because 
of the integral in the decay term in Eq.~\eqref{eq:integro_evol}.}
\label{fig:free}
\end{center}
\end{figure}

In Fig.~\ref{fig:free}, we represent the propagation of free particles in a 
BH chain consisiting of 100 sites within the $\mathcal{Q}$ region. 
The source is located at the first site in this region. 
The $\mathcal{L}$ and $\mathcal{R}$ regions are treated according 
to TBC, CAP or SECS. 
We chose the smoothing parameter $\lambda=0.1$ and the maximal 
scaling angle $\theta=1.5$. 
We can see that the three methods agree with each other and reproduce
the analytical value \eqref{eq:freen} of the on-site density within
the $\mathcal{Q}$ region. 
Moreover, the three methods seem to be stable for long propagation times, 
which allows us to study the steady state of this scattering process
with confidence. 
For a total propagation time of $Jt=250$, the SECS method is found to be
slightly faster than the CAP method.
We believe that this is due to CAP absorbing the outgoing flux less 
efficiently than SECS; hence, more sites in the leads contribute to 
the error computed by the adaptive RKF method and consequently more 
time steps have to be taken in order to reach the final time. 
On the other hand, the TBC method, while being exact from the 
formal point of view, is substantially slower (about 1000 times for 
this particular comparison) than the other two methods.
This can be explained by the fact that the numerical evaluation of the 
integral in the decay term is very costly and and has to be re-done
at any individual time step since this integral is a convolution of the 
memory kernel with the local history of the wavefunction.

\subsection{Free case with populated leads}
\label{subsec:TW}

Let us now study the influence of nonvanishing initial populations
in the leads. 
We generate the initial condition on the site $\ell$ according to
\begin{equation}
\label{eq:InitCond}
\psi_\ell(t=t_0) = \frac{1}{2}\left(\mathcal{A}_\ell + i \mathcal{B}_\ell\right),
\end{equation}
where $\mathcal{A}_\ell$ and $\mathcal{B}_\ell$ are real, independent 
Gaussian random variables with unit variance and zero mean, such that
\begin{subequations}
 \begin{align}
   \langle \mathcal{A}_\ell \rangle &=  \langle \mathcal{B}_\ell \rangle = 0,\\
   \langle \mathcal{A}_{\ell'}\mathcal{A}_\ell \rangle &=  \langle \mathcal{B}_{\ell'}\mathcal{B}_\ell \rangle = \delta_{\ell',\ell},	\\
   \langle \mathcal{A}_{\ell'}\mathcal{B}_\ell \rangle &= 0
 \end{align}
\end{subequations}
for all $\ell,\ell'\in \mathbb{Z}$.
As we shall point out in a forthcoming paper \cite{DujArgSch}, such 
initial conditions arise when applying the truncated Wigner method to 
the transport of Bose-Einstein condensates in the context of atom lasers. 

\begin{figure}[t]
\begin{center}
\input{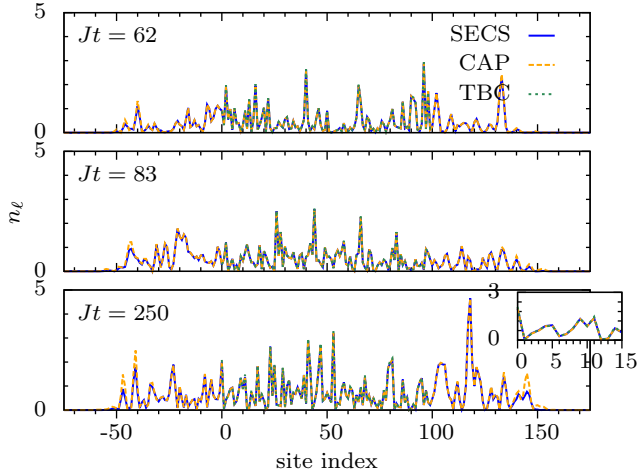}
\caption{(color online) Density profiles for free propagation 
(\textit{i.e.} $g_\ell=V_\ell=0$) with random initial populations
in the leads of the BH chain. 
There is no source of atoms in this calculation. 
As is clearly visible in the inset, all three methods compare 
very well in the central region $\mathcal{Q}$ which is again 
defined from site $1$ to site $100$.
SECS absorbs the outgoing flux more effectively in the leads than CAP
and appears to be the most efficient method in terms of computation time.}
\label{fig:noise}
\end{center}
\end{figure}

We consider again a homogeneous non-interacting BH chain and
used the same parameters as in Sec.~\ref{subsec:free} for SECS and CAP. 
Owing to the linearity of the time evolution in the case $g_\ell=0$ for all 
$\ell\in\mathbb{Z}$, we can, without loss of generality, set the coupling 
to the source to zero, $\kappa(t)=0$, for all times $t$ and study the
evolution of the random initial populations within an isolated waveguide (since
the effect of the source was already investigated in Sec.~\ref{subsec:free}).
The results of this simulation are shown in Fig.~\ref{fig:noise}. 
We arrive at the same conclusions as in Sec.~\ref{subsec:free}:
All three methods agree with each other and yield nearly identical 
on-site densities.
In particular, there is no artificial accumulation of the total population
within the central $\mathcal{Q}$ region due to an inefficient absorption of
the outgoing flux at the boundaries.
Again, the computational effort for SECS is appreciably lower than for 
CAP and substantially lower than for the integro-differential TBC method.

\subsection{Quantum dot}
\label{subsec:QD}

\begin{figure}[t]
\begin{center}
  \begin{psfrags}
    \psfrag{i}{{\small $\ell_0$}}
    \psfrag{i+6}{\hspace{-0.2cm}{\small $\ell_0+6$}}            
    \psfrag{lS}{{\small $\ell_S$}}
    \psfrag{Vl}{{\small $V_\ell$}}      
    \psfrag{V}{{\small $V$}}            
    \includegraphics[width=0.9\linewidth]{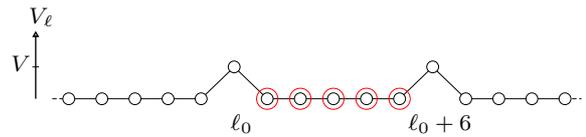}
  \end{psfrags}
\end{center}
\caption{(color online) One dimensional chain for the quantum dot model (see Eq.~\ref{eq:qdotV}). 
Plotted is the on-site potential $V_\ell$ as a function of the site 
index $\ell$.
The red circles represent sites where the atoms can interact. }
\label{fig:qdot_geom}
\end{figure}

In this section we study the effects of a nonvanishing potential and 
a finite on-site interaction on the scattering process.
As displayed in Fig.~\ref{fig:qdot_geom}, we specifically consider a 
double-barrier configuration defined by
\begin{equation}
V_\ell = V(\delta_{\ell,\ell_0} + \delta_{\ell,\ell_0+6}), \label{eq:qdotV}
\end{equation}
which can be seen as an atomic quantum dot.
Interaction is present only within the dot, \textit{i.e.} we define
\begin{equation}
  g_\ell = g \sum_{j=1}^{5}\delta_{\ell,\ell_0+j} \label{eq:qdotg}
\end{equation}
and choose $g = 0.1J$.

\begin{figure}[t]
\begin{center}
\input{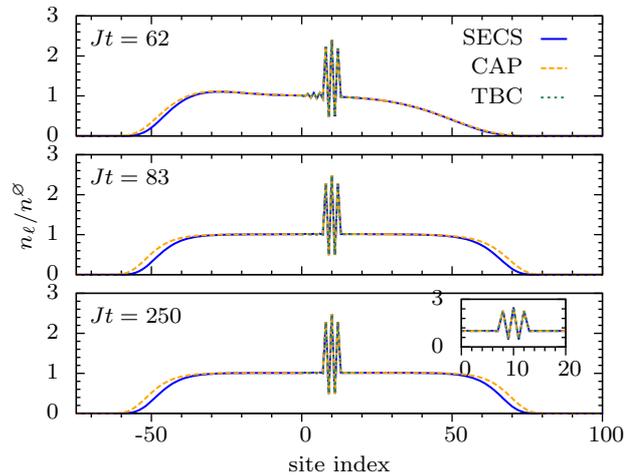}
\caption{(color online) Density profiles for a resonant propagation 
across a quantum dot in the BH chain, defined in Eqs.~\eqref{eq:qdotV} 
and \eqref{eq:qdotg}, with $\mu~=~-0.242J$ and $g~=~0.1J$ before (upper 
panel) and after (middle and lower panels) reaching the stationary state.
The source of atoms is located at site 1. 
As shown in the inset, all three methods compare very well in the 
central region $\mathcal{Q}$ defined from site $1$ to site $20$.}
\label{fig:qdot}
\end{center}
\end{figure}

\begin{figure}[t]
\begin{center}
\input{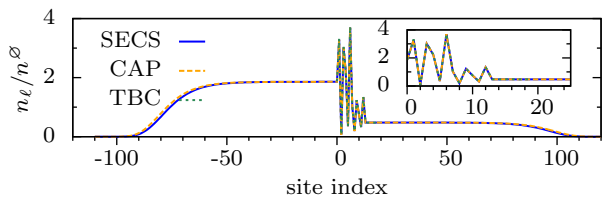}
\caption{(color online) 
Same as Fig.~\ref{fig:qdot} for $Jt = 250$ and the chemical potential 
$\mu~=~-0.8J$ which at $g~=~0.1J$ gives rise to non-resonant transport 
with finite reflection.
Again, very good agreement between all three methods is found.}
\label{fig:qdotlow}
\end{center}
\end{figure}

Figure \ref{fig:qdot} shows the density profiles obtained by the TBC, SECS, 
and CAP methods for the chemical potential $\mu=-0.242J$ which corresponds 
to a resonance of the double-barrier configuration at the interaction 
strength $g=0.1J$. 
Again, the three methods yield nearly identical results, which confirms 
their validity. 
This does not change if we add, as in Sec.~\ref{subsec:TW}, nonvanishing 
initial populations in the leads.
Finally, Fig.~\ref{fig:qdotlow} shows the case of imperfect transmission 
at the chemical potential $\mu=-0.8J$. 
As in the noninteracting cases studied in the previous subsections, 
we find that the SECS method is more efficient 
(\textit{i.e.} less time consuming) than the CAP and TBC methods.

\section{Conclusions}
\label{sec:Conclusion}

In summary, we introduced in this work the method of smooth exterior complex 
scaling (SECS) to the mean-field description of the one-dimensional transport 
of Bose-Einstein condensates within a guided atom laser configuration.
While this method is formally exact in a continuous system, imprecisions 
necessarily arise if the space is discretized in the framework of a 
finite-difference representation of the Gross-Pitaevskii equation.
We showed how to avoid this problem by choosing a sufficiently large 
smoothing parameter in the implementation of SECS.
A comparison with the (numerically inefficient) introduction of perfectly 
transparent boundary conditions, which are obtained from an analytical 
elimination of the semi-infinite leads of the waveguide (assuming that 
spatial inhomogeneities and nonvanishing interactions are restricted 
to a finite scattering region within the waveguide), yields very good 
quantitative agreement.
This was specifically tested for the case of resonant and non-resonant 
transport through an atomic quantum dot configuration consisting of a 
sequence of two symmetric barriers within which the interaction was 
assumed to be finite.
We furthermore showed that the SECS method is appreciably more efficient 
than the method of complex absorbing potentials that are defined with a 
comparable smoothing parameter.

In contrast to the method of absorbing boundary conditions proposed in 
Ref.~\cite{Shi91PRB}, which are adapted to outgoing waves with relatively 
well-defined wave numbers, the SECS method can also account for the presence 
of density fluctuations that arise from finite random initial populations 
within the waveguide of the atom laser.
This implies that SECS can be applied in the framework of the Truncated 
Wigner method \cite{Steel1998} which approximately accounts for the effect 
of quantum fluctuations beyond the mean-field description of the propagating 
condensate.
This specific application, as well as the use of SECS within the many-body
matrix product state (MPS) algorithm \cite{Verstraete2004,Vidal2003,Vid04PRL}, 
shall be discussed in a forthcoming publication \cite{DujArgSch}.

\section*{Acknowledgements}

We acknowledge financial support from Fonds de la Recherche Scientifique de Belgique (F.R.S.-FNRS) and the DFG Forschergruppe FOR760 "Scattering Systems with Complex Dynamics". The computational resources have been provided by the Consortium des \'{E}quipements de Calcul Intensif (C\'{E}CI), funded by the F.R.S.-FNRS under Grant No. 2.5020.11.


\end{document}